\documentclass[twocolumn,english,prb,aps,10pt]{revtex4-2}
\usepackage{algpseudocode}
\usepackage{amsmath}
\usepackage{amssymb}
\usepackage{babel}
\usepackage{color}
\usepackage{graphicx}
\usepackage{stmaryrd}       

\newcommand\bigO{\mathcal{O}}%
\newcommand\cee{\hat{c}^{\mathchoice{\vphantom{\dagger}}{}{}{}}}%
\newcommand\cdag{\hat{c}^{\dagger}}%
\newcommand\popcount{h}%
\newcommand\bitand{\ensuremath{\textsc{and}}}%
\newcommand\bitor{\ensuremath{\textsc{or}}}%
\newcommand\bitnot{\ensuremath{\textsc{not}}}%
\newcommand\bitxor{\ensuremath{\textsc{xor}}}%
\begin{document}
\title{Trie-based ranking of quantum many-body states}
\author{Markus Wallerberger}
\affiliation{Department of Solid State Physics, TU Wien, 1040 Vienna, Austria}
\author{Karsten Held}
\affiliation{Department of Solid State Physics, TU Wien, 1040 Vienna, Austria}
\begin{abstract}
Ranking bit patterns---finding the index of a given pattern in an
ordered sequence---is a major bottleneck scaling up numerical
quantum many-body calculations, as fermionic and hard-core bosonic states translate naturally
to bit patterns. Traditionally, ranking is done by bisectioning search, which
has poor cache performance on modern machines. We instead propose
to use tries (prefix trees), thereby achieving a two- to ten-fold
speed-up in numerical experiments with only moderate memory overhead.
For the important problem of ranking permutations, the corresponding
tries can be compressed. These compressed ``staggered'' lookups allow for a considerable
speed-up while retaining the memory requirements of prior algorithms
based on the combinatorial number system.
\end{abstract}
\maketitle

\section{Introduction}
At their first encounter with the quantum many-fermion problem, students
are usually warned that any brute-force attempt at a solution is destined
to shatter at the \emph{exponential wall}---the doubling of computing
requirements with the addition of each state. Commonly, this then
segues into the presentation of some polynomial-cost approximation.

So it is perhaps ironic that the brute-force solution of small many-fermion
problems, known as exact diagonalization (ED)~\cite{Bonner64,Alvermann2011,Innerberger2020}, many-body Lanczos
method~\cite{Lanczos1950,Cullum1985}, or full-configuration interaction (FullCI)~\cite{Siegbahn84,Knowles84},
has remained one of the workhorses of many-body physics and quantum
chemistry. On small systems, it serves as benchmark for more
advanced (classical~\cite{Motta17} or quantum computing~\cite{Gheorghiu19R}) methods
and to explore emergent many-body behavior~\cite{Sandvik10ACP}. For realistic
systems, exact diagonalization is particularly useful as kernel of
embedding theories~\cite{DMFT,DMET,SEET}, where one approximately maps the full
problem onto one or more difficult but small many-body problems. In
that problem space, ED is competing with Monte Carlo methods~\cite{Gull2011,w2dynamics},
which do not have as severe a restriction in system size, but usually
require sufficient symmetry to avoid a prohibitive sign problem.
An iterative ED is also at the core of more sophisticated renormalization schemes, most notably the numerical renormalization group (NRG)~\cite{Bulla2008} and density matrix renormalization group (DMRG)~\cite{Schollwock2005}  method.

\begin{figure}[t]
\centering\includegraphics[scale=.9]{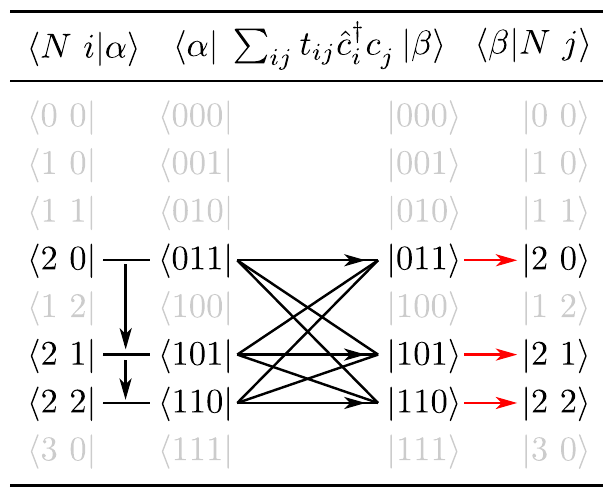}
\caption{On-the-fly construction of the matrix elements of a
  non-interacting three-site Hamiltonian in the symmetry sector $\hat N=2$:
  unranking the states into occupation number states $|\alpha=n_2n_1n_0\rangle$
  (left column), followed by applying  the Hamiltonian (center column)
  and ``ranking'', i.e., mapping back to the state index in the symmetry sector (right column).}
\label{fig:overview}
\end{figure}

Since scaling with system size is the fundamental limitation of ED,
we are seeking ways to mitigate it while retaining an unbiased deterministic
approach. Here, the Lanczos method combined with on-the-fly representation
of the Hamiltonian~\cite{Siegbahn84,Gagliano86} and use of quantum numbers~\cite{Bonner64} is the state-of-the-art in diagonalizing spin systems~\cite{Lin90SublatticeCoding} and gaining traction for
 many-fermion solvers~\cite{HPhi,libcommute} (see Fig.~\ref{fig:overview} and Sec.~\ref{sec:ed} for a recap). This
approach is still memory-bound, as the subspaces still grow exponentially,
albeit somewhat more slowly. The main runtime bottleneck is, perhaps
surprisingly, not the application of the Hamiltonian itself, but the
mapping of the many-body states back into the block structure generated
by the quantum numbers, a procedure known as ``hashing'' in the
ED community~\cite{Gagliano86} and ``ranking'' in computer science~\cite{Knuth4A}
(red arrows in Fig.~\ref{fig:overview}).
Aside from generic techniques for the mapping, such as bisection search
and hash tables, for special sectors explicit formulas have been put
forward for computing the rank by examining the position of each bit
in the state~\cite{Liang95} (see Sec.~\ref{sec:naive}).

In this paper, we first show how to create a set of small precomputed
tables for these explicit formulas, allowing us to process data in
chunks of multiple bits rather than one bit at a time at a considerable
speedup (see Sec.~\ref{sec:staggered}). We then generalize these
formulas to an arbitrary set of quantum numbers by leveraging a special
kind of $n$-any search tree called ``trie'' (Sec.~\ref{sec:trie})~\cite{Briandais59,Knuth3}.
Numerical experiments in Sec.~\ref{sec:exp} show significant speedups
for both microbenchmarks and real-world ground state computations.

\section{The finite many-fermion problem\label{sec:ed}}

To set the stage, let us briefly review the challenge of finding the
ground state of a system of interacting fermions in $M$ spinorbitals.
We need to find the lowest eigenvalue and associated eigenvector of
the Hamiltonian, a $2^{M}\times2^{M}$ matrix given in 2nd quantization by:
\begin{equation}
\hat{H}=-\sum_{i,j=0}^{M-1}t_{ij}\cdag_{i}\cee_{j}+\frac{1}{4}\sum_{i,j,k,l=0}^{M-1}U_{ijkl}\cdag_{i}\cdag_{j}\cee_{l}\cee_{k},\label{eq:H}
\end{equation}
where $i$, $j$, $k$, $l$ are spinorbital indices, $i=0, \ldots, M-1$, $t_{ij}$ are
hopping amplitudes, $U_{ijkl}$ are two-body interaction strengths,
$\cee_{i}$ is (a matrix) annihilating a fermion in spinorbital $i$,
and $\cee_i{}^\dagger$ is its Hermitian conjugate. The chemical
potential, if present, can be absorbed into the diagonal entries of
$t$.

The explicit form of $\hat{H}$ is easily constructed in the occupation
number basis. There, each basis state is one possible combination
of occupations $n_{i}$ of the spinorbitals, $|n_{M-1}\ldots n_{1}n_{0}\rangle$.
Since  for fermions $n_{i}\in\{0,1\}$,
each state is nothing
but a bit pattern. In order to form a matrix, we assume these patterns
are ordered, or ``ranked'', lexicographically. Interpreting a bit
pattern as a number in base two then gives us a natural correspondence
of a state and its rank $\alpha\in\{0,\ldots,2^{M}-1\}$ in the basis.
For example, the state $|011101\rangle$ has rank $\alpha=011101_{2}=29$.
The $i$-th annihilator is then a matrix confined to the $(2^{i})$'th
side diagonal:
\begin{equation}
\cee_{i}|n_{M-1}\ldots n_{i}\ldots n_{0}\rangle=\delta_{n_{i}1}\prod_{j<i}(-1)^{n_{j}}|n_{M-1}\ldots0\ldots n_{0}\rangle,\label{eq:cee}
\end{equation}
where $\delta_{ij}$ is the Kronecker delta and the product in the prefactor ensures
anticommutativity.

In implementing Eqs.~(\ref{eq:H}) and (\ref{eq:cee}), the motivated
but unseasoned physicist has to brace for a series of increasingly
painful concessions. First, at around $M=14$, the cost $\bigO(2^{3M})$
of numerical diagonalization becomes prohibitive and Krylov subspace
methods can be used instead~\cite{Lanczos1950,Cullum1985}. Secondly, at $M\sim18$, the
memory $\mathcal{O}(2^{2M})$ required to store $H$ starts to blow
up, and one may switch to sparse storage~\cite{EDLib} as it also combines
neatly with Krylov methods: there, we do not need  to construct $H$ explicitly,
but only its applications on vectors $H|\psi\rangle$. Sparse storage
requires $\mathcal{O}(K'2^{M})$ memory, where $K'$ is the number
of unique side diagonals created by Eq.~(\ref{eq:cee}), which becomes
problematic for $M\sim22$. One may then try to compress multiple
columns into bitsets~\cite{EDLib}, which will get one to $M\sim26$. Finally one
may use massive parallelization to distribute the required memory~\cite{Jia18,Lauchli18},
which for a supercomputer of reasonable size will break down at $M\sim32$.
(These limits vary with system type, technical prowess of the implementer
and, assuming ``Moore's law'' holds, should be incremented by one
every 18 months or so.)

The key to compressing $H$ further is to realize that Eq.~(\ref{eq:H})
already constitutes a highly compressed form: a sum of $\mathcal{O}(M^{4})$
terms, each of which is a product of creation and annihilation operators with a
scalar prefactor.  For each term $T$ of this form, there exists a tuple $(m,r,x,s,v)$
such that the application on any occupation basis state $|\alpha\rangle$ is given by
the following, extremely efficient formula~\cite{Siegbahn84,Gagliano86}:
\begin{equation}
T|\alpha\rangle=v\delta_{\alpha\owedge m,r}(-1)^{\popcount(\alpha\owedge s)}|\alpha\oplus x\rangle.\label{eq:Talpha}
\end{equation}
Here, $\owedge$ denotes bitwise \bitand, $\oplus$ denotes bitwise
\bitxor, and $h(x)$ is the Hamming weight---the number of set bits---of $x$.
(For completeness, we show how to construct these tuples in Appendix~\ref{app:mul}.)
Eq.~(\ref{eq:Talpha})
allows one to compute $H|\psi\rangle$ in $\bigO(K2^{M})$ time, where
$K$ is the number of terms in $H$. $K\ge K'$ as $K'$ is equal to
the number of unique values of $x$, so we increase runtime, but require only $\bigO(K)$ memory for storing
the tuples for each term in $H$.

This takes care of applying operators, however, eventually the memory
$\mathcal{O}(2^{M})$ required to store a single vector will become
a problem. To mitigate this, we can use the fact that the Hamiltonian
(\ref{eq:H}) conserves particle number (commutates with the particle number operator  $\hat{N}$):
\begin{equation}
[\hat{H},\hat{N}]\equiv \sum_{i=0}^{M-1}[\hat{H},\cdag_{i}\cee_{i}]=0,\label{eq:HNcomm}
\end{equation}
which partitions the Hamiltonian into $M+1$ blocks with different particle numbers ${N}=0,1,\ldots,M$.
Since   $\hat{N}$ is diagonal in the occupation number basis, each
state $|\alpha\rangle$ can be assigned to exactly one block. For
this, we introduce the notation $|\alpha\rangle\equiv|N\,i\rangle$,
where $N$ is the block number (the number of set bits in $\alpha)$
and $i$ is the rank of $\alpha$ within its block. As with the full
space, we choose $|\alpha\rangle$ to have rank $i$ if it is the
$i$-th state, ordered lexicographically within the $N$-th block.

Since $\hat{H}$ is now block-diagonal thanks to particle number conservation, we can treat each block separately
and then collect the results afterwards. This only involves states in  the
matrix--vector product confined to one block, i.e.:
\begin{equation}
\langle N\,i|\hat{H}|\psi\rangle=\sum_{j=0}^{\binom M N -1}\langle N\,i|\hat{H}|N\,j\rangle\langle N\,j|\psi\rangle.\label{eq:mvblock}
\end{equation}
The number of states in the block $N$ is equivalent to the
number of choices of $N$ set bits out of $M$, so the storage requirements
for a block vector drop to $\binom M N$. Since the largest block
is half-filled ($N=M/2$) and
\begin{equation}
\binom M {M/2}\approx\left(\frac{\pi}{2}M\right)^{-\frac{1}{2}}2^{M}\label{eq:asympN}
\end{equation}
for large $M$, we do not substantially affect the scaling, but the
prefactor allows us to go to slightly larger systems, $M\to M+3$ or so.

Combining Eq.~(\ref{eq:mvblock}) with on-the-fly application (\ref{eq:Talpha})
involves yet another complication, which shall become the focus of
this paper: in order to apply some term $T$, we first need to \emph{unrank}
$|N\,j\rangle$, i.e., map it to its corresponding $|\alpha\rangle$ for evaluating  Eq.~(\ref{eq:Talpha}),
and after we have computed $|\beta\rangle=T|\alpha\rangle$, we need
to \emph{rank} it, i.e., map $|\beta\rangle$  back to its $|N\,i\rangle$ (cf.~Fig.~\ref{fig:overview}).
Unranking is the easier of the two problems, as we simply need to
maintain a list of $|\alpha\rangle$ for each $|N\,i\rangle$. Ranking
is trickier to do quickly: we cannot simply use a lookup table which
maps $|\alpha\rangle$ to $|N\,j\rangle$, as this would again require
$\mathcal{O}(2^{M})$ memory. Hithero, one typically uses bisectioning search
into the list of many-body states, which is slow on modern machines.
As a result, it is ranking and not Eq.~(\ref{eq:Talpha}) that is
usually the bottleneck of finite many-body calculations.

\section{Combination ranking\label{sec:combrank}}

As outlined in the previous section, we are concerned with ranking and unranking
the $\binom M N$ basis vectors having $N$ electrons for $M$ spinorbit states (or, in general, bitpatterns of length $M$ with $N$ bits set to 1).
The set of all  such \emph{combinations}~ $C_{N}^{M}$~\cite{Ryser63} can be written as
\begin{equation}
C_{N}^{M}=\{(c_{N},\ldots,c_{1}):M>c_{N}>\cdots>c_{1}\ge0\},\label{eq:comb}
\end{equation}
where $c_i$  conveys the $i$th selected elements out of the $M$ possibilities. The task is now to assign a  \emph{rank} $i\in\{0,\ldots,{\binom M N}-1\}$ to each
element of $C_{N}^{M}$. We do so lexicographically,
i.e., first order  by $c_{N}$, then by $c_{N-1}$, and so forth till  $c_{1}$.
For example, the set $C_{3}^{4}$ contains the elements $(2,1,0),(3,1,0),(3,2,0),$
and (3,2,1)  which are thus assigned the  lexicographical ranks
$0,1,2$ and 3, respectively.

\subsection{Ranking using combinadics\label{sec:naive}}

A convenient way of computing this rank is the \emph{combinatorial
  number system} or \emph{combinadics}~\citep{Pascal1887,Knuth4A}
which allows one to compute the lexical rank  by calculating
the number of combinations ordered before
the current one as~\citep{Pascal1887,Knuth4A}:
\begin{equation}
(c_{N}\ldots c_{1})_{C}:={\binom {c_{N}} N}+\cdots+{\binom {c_{2}} 2}+{\binom {c_{1}} 1},\label{eq:cns}
\end{equation}
where  $\binom c n=0$ whenever
$n>c$. We can comprehend Eq.~(\ref{eq:cns}) by realizing that there are
$\binom {c_{N}} N$ possibilities to select $N$  elements with  flavors
$0 \ldots c_N-1$ that  have lexicographically the leading ($N$th) bit set to 1 before
$c_{N}$; then for the $(N-1)$th element  there are
$\binom{c_{N-1}}{N-1}$ such possibilities
and so forth.
For example, $310_{C}= \binom 3 3 = 1$, which reflects that  $(3,1,0)$ is ranked second
after $(2,1,0)$ in $C_{3}^{N}$.

We contrast this with the ``bit pattern''
of each element:
\begin{equation}
\alpha(c_{N},\ldots,c_{1})=2^{c_{N}}+\cdots+2^{c_{1}}\label{eq:alphans}
\end{equation}
For the same example, we have $\alpha(310)=1011_{2}=11$.

We can
thus define a rank and unrank function by the following identities:
\begin{align}
\mathrm{rank}(\alpha(c_{N},\ldots,c_{1})) & =(c_{N},\ldots,c_{1})_{C},\label{eq:rank}\\
\mathrm{unrank}((c_{N},\ldots,c_{1})_{C}) & =\alpha(c_{N},\ldots,c_{1}).\label{eq:unrank}
\end{align}

\begin{figure}
\begin{algorithmic}[1]
\Function{{\rm $i = $} rank}{$\alpha$}
    \State $i \gets 0$
    \State $k \gets 1, c \gets 0$
    \While{$\alpha \neq 0$}
        \State $c \gets \operatorname{ctz}(\alpha)$
        \State $\alpha \gets \alpha \owedge (\alpha - 1)$
        \State $i \gets i + \binom c k$
        \State $k \gets k + 1$
    \EndWhile
\EndFunction
\end{algorithmic}

\caption{Combinadics algorithm  for ranking a bit pattern using  Eq.~(\ref{eq:cns}) \citep{Liang95}.
Here,  $\operatorname{ctz}(x)$
counts the number of trailing zeros in the binary representation of
$x$,  $\owedge$ denotes bitwise \bitand, and $\alpha \owedge (\alpha - 1)$ is an efficient means to remove the rightmost bitwise 1 from  $\alpha$. The set of binomial coefficients $\binom c k$ should be
precomputed and stored.}
\label{fig:algr}
\end{figure}

Eq.~(\ref{eq:cns}) allows one to  compute $\mathrm{rank}(\alpha)$,
known also as ``perfect hashing'' in the context of exact diagonalization
\citep{Liang95,Jia18}. We reproduce this algorithm in Figure~\ref{fig:algr}.
The basic idea is to extract the position $c_{k}$ of the $k$-th
set bit by using the count trailing zeros (ctz) instruction, available
on most modern CPUs, followed by clearing the least significant bit.
One then adds the corresponding binomial coefficient  of Eq.~(\ref{eq:cns}) to the rank $i$.
These coefficients should be precomputed for all $0\le c<M$ and $1<k\le M$,
which comes at a moderate memory cost of $32\thinspace\mathrm{KiB}$ for $M=64$ (64-bit numbers).

A priori it is not clear why this algorithm should be faster than
simply maintaining an ordered list of all elements of $C_{N}^{M}$
and finding a representative by bisectioning, since from Eq.~(\ref{eq:asympN})
we expect $\bigO(M)$ steps are needed for both algorithms in the
half-filled case. However, depending on the exact structure of the
Hamiltonian, ranking by Eq.~(\ref{eq:cns}) can be significantly
faster on modern machines, because: (i) bisectioning heavily relies
on efficient branch prediction, as at each step we have to choose
which ``half'' of the list we will focus on; and (ii) bisectioning
has poor cache locality, as we have to jump around the complete list.
In contrast to that, the algorithm in Figure~\ref{fig:algr} is essentially
branch-free except for the loop condition, and the lookup table is
small enough such that the CPU may reasonably keep it in its cache.

\begin{figure}
\begin{algorithmic}[1]
\Function{{\rm $i = $} rank-fast}{$\alpha$}
    \State $i \gets 0$
    \State $N' \gets 0, M' \gets 0$
	\While{$\alpha \neq 0$}
        \State $r \gets \alpha \bmod 2^R$
        \State $i \gets i + \mathrm{rank}(r, M', N')$
        \State $M' \gets M' + R$
        \State $N' \gets N' + h(r)$
        \State $\alpha \gets \lfloor \alpha / 2^R \rfloor$
    \EndWhile
\EndFunction
\end{algorithmic}

\caption{Improved combinadics algorithm
with staggered lookup,  using Eq.~(\ref{eq:partrank}) for ranking. Here, $\bmod$
denotes the binary modulo operation and $h(x)$ is the Hamming weight,
i.e., the number of set bits, in $x$. The values $\mathrm{rank}(r,M',N')$
should be precomputed and stored.
}
\label{fig:fastrank}
\end{figure}

For unranking, Eq.~(\ref{eq:unrank}), we only have a (relatively)
small number of $\binom M N$ elements  for each  $C_{N}^{M}$.
For these the bitpattern $\alpha$  can be stored in a lookup table.

\subsection{Staggered lookup\label{sec:staggered}}

While
a full lookup table for $\mathrm{rank}(\alpha)$ is prohibited by
the memory cost,
we can further enhance the numerical efficiency of the ranking algorithm Figure~\ref{fig:algr}  by splitting
up $\alpha$ into chunks of $R$ bits, starting from the least significant
ones, and employ for these chunks of bits
precalculated ranks. This  balances computational cost vs.~memory cost and can be optimized with respect to $R$.

If in the previous chunks of $\alpha$, we already encountered  $M'$ succeeding flavors and $N'$ succeeding occupations,
the next chunk of bits with occupations $(c_{Q},\ldots,c_{1})$ adds the following binominals to
Eq.~(\ref{eq:cns}):
\begin{equation}
\begin{split} & \mathrm{rank}(\alpha(c_{Q},\ldots,c_{1}),M',N')\\
 & \quad:=\binom{M'+c_{Q}}{N'+Q}+\cdots+\binom{M'+c_{2}}{N'+2}+\binom{M'+c_{1}}{N'+1},
\end{split}
\label{eq:partrank}
\end{equation}
In the corresponding  algorithm  Figure~\ref{fig:fastrank},
we
keep track of
how many bits $M'$ and how many set bits $N'$ we have encountered
previously. We  then use Eq.~(\ref{eq:partrank}) for the next chunk of $R$ bits of  $\alpha$, remove these bits, and update $M'$ and $N'$ before turning to the next  bits.

Crucially, $\mathrm{rank}(\alpha,M',N')$, unlike $\mathrm{rank}(\alpha)$,
can be precomputed and stored provided $R$ is not too large. For
a given bound $M$, one needs to precompute for $M'\in\{0,R,2R,\ldots,M-R\}$.
For a given $M'$, one needs to precompute for $N'=\{0,1,\ldots,M'\}$
and all possible values of $\alpha\in\{0,1,\ldots,2^{R}-1\}$. This
gives a total memory demand of:
\begin{equation}
2^{R}\sum_{M'}(M'+1)=2^{R}(M-R+2)\left(\frac{M}{2R}+1\right)\label{eq:stagmem}
\end{equation}
numbers. For $M=64$ and $R=8$, this requires 580~KiB of memory,
and the corresponding lookup table can be held in cache on reasonably
modern machines.

Let us conclude with a couple of remarks on the lookup table: firstly,
a table constructed in aforementioned fashion is universal in the
sense that it works for any sector; we can save memory by restricting
ourselves to the values of $N'$ to those admittable for a given $N$
at the expense of a slightly more complicated lookup logic. Since
the table is of moderate size, it is usually not necessary to do so.
However, the table should be stored with the $\alpha$ index varying
fastest to ensure that only those values of $(M',N')$ that are actually
used are loaded as cache lines. Secondly, in computing the entries
of the lookup table, the algorithm in Figure~\ref{fig:algr} can
be reused, since:
\begin{equation}
\begin{split} & \mathrm{rank}(\alpha(c_{Q},\ldots,c_{2},c_{1}),M',N')\\
 & \quad=\mathrm{rank}(\alpha(M'+c_{Q},\ldots,M'+c_{2},M'+c_{1},\\
 & \qquad\qquad\qquad N'-1,\ldots,1,0)).
\end{split}
\label{eq:reuse}
\end{equation}
Thirdly, one may not only store $\mathrm{rank}(\alpha,M',N')$ for
a given $\alpha$, but also the offset in the lookup table corresponding
to the updated values of $M'$ and $N'$. This saves the computation
of the Hamming weight and some index manipulation, yet doubles the
memory demand. We empirically find this to be a beneficial tradeoff
on most modern CPUs, and have employed it in all benchmarks.

Similar to the standard combination ranking algorithm (Figure~\ref{fig:algr}),
the fast ranking algorithm can be made essentially branch-free. (The ``while'' loop can turned into a
for loop if $M$ is added as an argument, and unrolled if $M$ is known a priori.) Unlike
the standard algorithm, only $\lceil M/R\rceil$ instead of $N$ lookups
are required, as $R$ bits are processed at a time. The rest of the
manipulations are cheap, since the computation of the Hamming weight
is available as a separate CPU instruction on all common machines.
For the ``bottle-neck'' case of $N\approx M/2$ and the choice $R=8$,
we thus expect a significant speedup. This is evidenced  numerically
in Sec.~\ref{sec:exp}.

\section{Trie-based ranking\label{sec:trie}}

If our Hamiltonian (\ref{eq:H}) only conserves particle number,
we can use the algorithm presented in Sec.~\ref{sec:staggered} to
rank. Commonly, however, the Hamiltonian will have more symmetries.
For instance, if $H$ conserves total spin as well, it conserves both
the number $N_{\uparrow}$ and $N_{\downarrow}$ of spin-up and spin-down
particles, respectively. Assuming that the least significant bits
correspond to spin-down, we have:
\begin{equation}
\mathrm{rank}_{N,S_{z}}(\alpha)=\mathrm{rank}_{N}(\alpha_{\uparrow})
\binom{M/2}{N_{\uparrow}}+\mathrm{rank}_{N}(\alpha_{\downarrow}),\label{eq:rankupdown}
\end{equation}
where $\alpha_{\uparrow}=\lfloor\alpha/2^{M/2}\rfloor$ and $\alpha_{\downarrow}=\alpha-\alpha_{\uparrow}$;  $\mathrm{rank}_{N}$ is given by Eq.~(\ref{eq:rank}) as before, allowing us
to reuse the corresponding fast algorithm (Fig.~\ref{fig:fastrank}).

For other sets of quantum numbers, e.g., total momentum or orbital
parity~\cite{Parragh12PS}, the situation is different: there, we do not usually have
an  expression that is both, compact and fast, for ranking states from the corresponding
sector, and need to fall back on a search. Ideally, we would like
to carry over the advantages from the staggered lookup in a combination
to this case: a fast, cache-friendly search.

\begin{figure}
  \includegraphics[width=1\columnwidth]{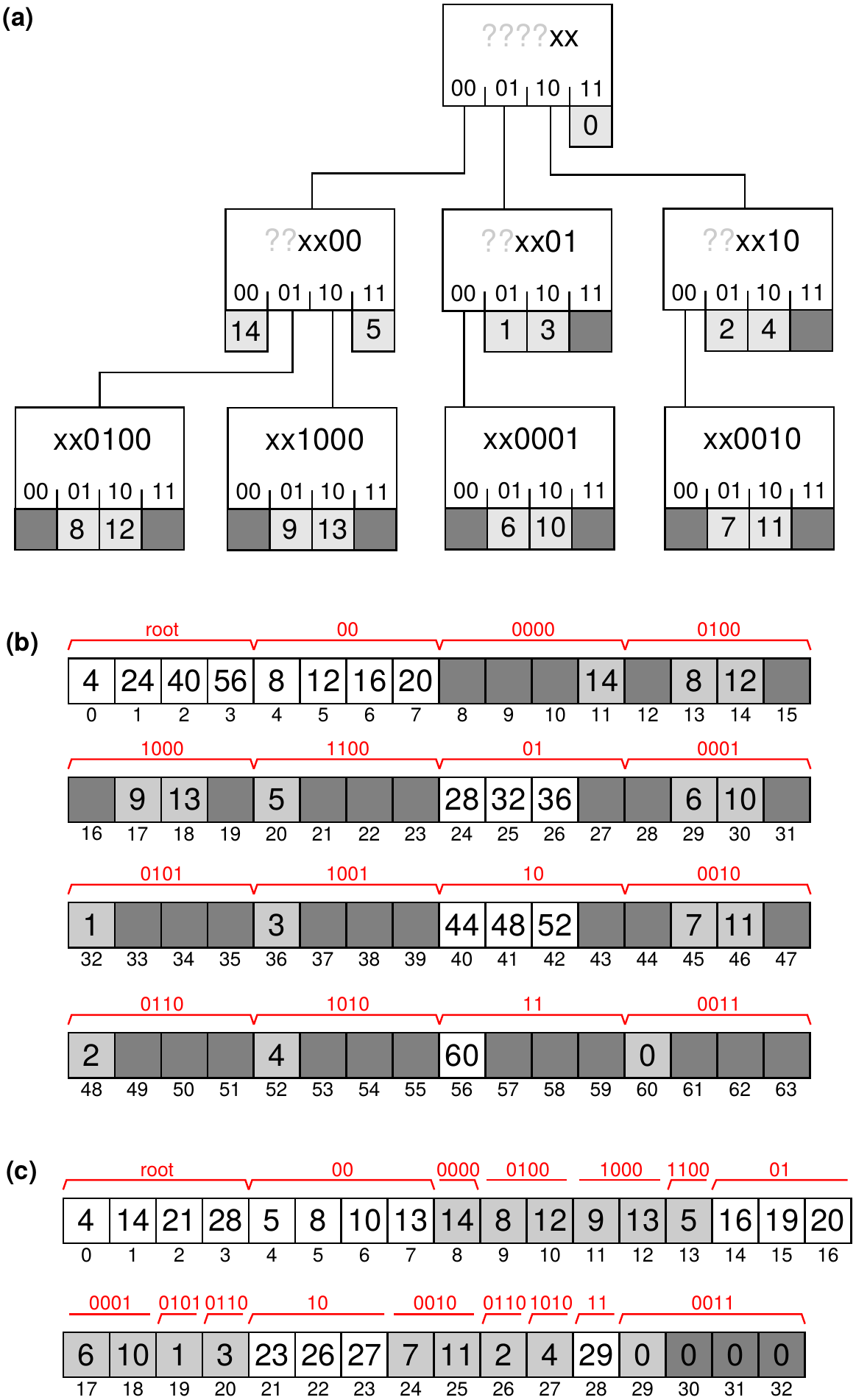}
  \caption{Trie algorithm for ranking, exemplified for $M=6$, $N=2$ and $R=2$.
    Here, we dispatch
the least significant two bits, branch to the next node according to these, then shift and repeat. (a)
Full trie, where white blocks represent branch nodes, grey boxes with numbers
represent leaves. Suffix compression \cite{Liang83PhD} is employed to reduce the trie
size. (b) Linearized representation of the same trie for saving memory. (c) Packed
representation to further reduce memory.}

\label{fig:trie}
\end{figure}

To this end, let us understand the fast ranking algorithm Fig.~\ref{fig:fastrank} as lookup
in a tree index. For the case of $N=2$ particles in $M=6$ flavors
and a chunk size of $R=2$ this is depicted in Fig. \ref{fig:trie}a:
for any pattern $\alpha$, we start at the root node. We dispatch
on the $R$ least significant bits $(\alpha_{1}\alpha_{0})$ and follow
the $(\alpha_{1}\alpha_{0})$  branch  to the next node. We then shift $\alpha$ by
2 bits to the right and repeat the procedure until we arrive at a
leaf node, which contains the rank.

This procedure already suggests a generalization of the lookup for
arbitrary quantum number sets: Elevating Fig.~\ref{fig:trie}a to
a data structure, we have constructed a \emph{trie }(or prefix tree)~\cite{Briandais59,Knuth3}.
A trie of radix $2^R$ for bit patterns of length $M$ is a search tree
of height $\lceil M/R\rceil$ with a branching factor of $2^{R}$,
i.e., each non-leaf node of the trie maintains an array of size $2^{R}$,
which are pointers to descendant nodes. The leaf nodes instead represent
the index $i$. (This is not exactly congruent with Fig.~\ref{fig:trie}a,
where we have omitted branches with only one possible path, a procedure
known as suffix compression.) Tries maintain two important benefits
of the staggered lookup algorithm: they process data in chunks of
$R$ bits, thus only requiring $\lceil M/R\rceil$ lookups, and they
replace branching by an indexing operation into an array of size $2^{R}$.

Tries, however, are not ideal in terms of memory locality and require
significant memory overhead. To mitigate this, we can switch to a linearized
representation~\cite{Liang83PhD}, depicted in Fig.~\ref{fig:trie}b: the trie
is represented by a single contiguous array $(t_{0}\ldots t_{T-1})$,
which is efficient since the trie is not mutated once created. Each
node is then represented by an index $k$ in the array (the root node
has index 0). For a branch node, the element $t_{k+r}$ contains the
index of the child for the branch corresponding the $R$ active bits
of the state being equal to $r$ (white boxes). For a leaf node, $t_{k}$
corresponds to the rank of the state (gray boxes).

We note that this structure is compatible with \emph{packing}~\cite{Liang83PhD}:
e.g., in Fig.~\ref{fig:trie}, the branch node corresponding to $10_2$
does not have a child for $r=11_2$. We can thus omit the element for
$r=11_2=3$. In general, a sequence of forbidden trailing descendants can
be omitted. The same is true for forbidden descendants at the beginning:
there, we omit the elements and move the index in the parent node
forward by the number of elements deleted. For example, the node 0000
only allows $r=11_2$, so we can omit the first three elements. Consequently
the index $t_{4}$, which is the corresponding element in node $00_2$,
contains 5 rather than 8, reflecting the omission of 3 elements. As illustrated in
Fig.~\ref{fig:trie}c, packing thus significantly reduces the memory demand in cases where the trie is only sparsely populated. (Even more advanced packing strategies, where ``holes'' in the middle
of the index are kept track of and filled by suitable nodes, are possible but beyond the
scope of this paper.)

\begin{figure}
\begin{algorithmic}[1]
\Function{{\rm $i = $} rank-trie}{$t, \alpha$}
    \State $i \gets 0$
    \For{$i \in \{1, 2, \ldots, \lceil M/R\rceil - 1\}$}
        \State $r \gets \alpha \bmod 2^R$
        \State $i \gets t_{i+r}$
        \State $\alpha \gets \lfloor\alpha / 2^R\rfloor$
    \EndFor
    \State $i \gets t_{i+\alpha}$
\EndFunction
\end{algorithmic}

\caption{Algorithm for ranking a state using linearized tries. Here $\bmod$ denotes the
binary modulo operation; Fig.~\ref{fig:trie} exemplifies $t_i$.}
\label{fig:trielookup}
\end{figure}

Fig.~\ref{fig:trielookup} presents the algorithm for ranking a state
using the linearized (and optionally packed) trie. Let us walk through
the algorithm for, e.g., the trie in Fig.~\ref{fig:trie} and $\alpha=100100_{2}$:
we start at the root with $i=0$. Since $r=\alpha\owedge11_{2}=00_{2}=0$
and $t_{0}=4$, we move to $i=4$ and consider the next two bits of
the state, $r=01_{2}$. Note that $t_{5}=8$, even though the data
for node 0100 starts at index 9, as we have omitted the unused first
child. Finally, we have $\alpha=10_{2}$ and the rank is given as
$t_{8+2}=12$.

As with staggered lookup, we can now trade off memory overhead with
lookup performance by adjusting the radix $R$. Unlike staggered lookup,
the memory overhead now scales with the number $N_{s}$ of states
in the sector, typically requiring $\bigO(fN_{s})$ space, where the
overhead factor $f\ge1$ depends on $R$ and the structure of the
sector. In the example above, we have $f\approx2$.

Let us add a remark on the trailing elements in the packed
trie (Fig.~\ref{fig:trie}c): we could omit those, since they are
not pointing to any valid rank. However, including the full root node
and the trailing elements of the last leaf node and setting the unused
indices to zero ensures that the lookup procedure (Lines~5 and 8
in Fig.~\ref{fig:trielookup}) will only ever access valid array
indices, regardless of whether $\alpha$ is a valid state or not.
This allows us to modify the algorithm in Fig.~\ref{fig:trielookup} to include a
cheap check whether $\alpha$ is indeed a valid state in the sector:
after the lookup, we simply verify that $t_{i}=\alpha$ by consulting the unrank lookup table.

Finally, let us turn  back to our initial consideration, i.e., making the ranking algorithm suitable to conservation laws beyond the total number of particles (per spin sector). This is possible by simply eliminating all leaves and branches that do not fulfill a given conservation law.  In practice, one starts by generating all states on the fly which fulfill particle number conservation in lexicographic order.  Next, one filters out those states swhich do not satisfy additional conservation law.  This iterator over state numbers is then fed into the trie generation routine, which can directly generate the packed trie.

\section{Numerical experiments\label{sec:exp}}

\subsection{Ranking microbenchmark\label{sec:micro}}

To compare the established methods against one another, we first perform
a microbenchmark focusing on lookup performance alone. For this, we
consider $M=28$ spinorbitals and conserved total occupation $\hat{N}$
and analyze all sectors $N=1,\ldots,27$, which corresponds to ranking
the combination set $C_{N}^{28}$.

\begin{figure}
\includegraphics[width=1\columnwidth]{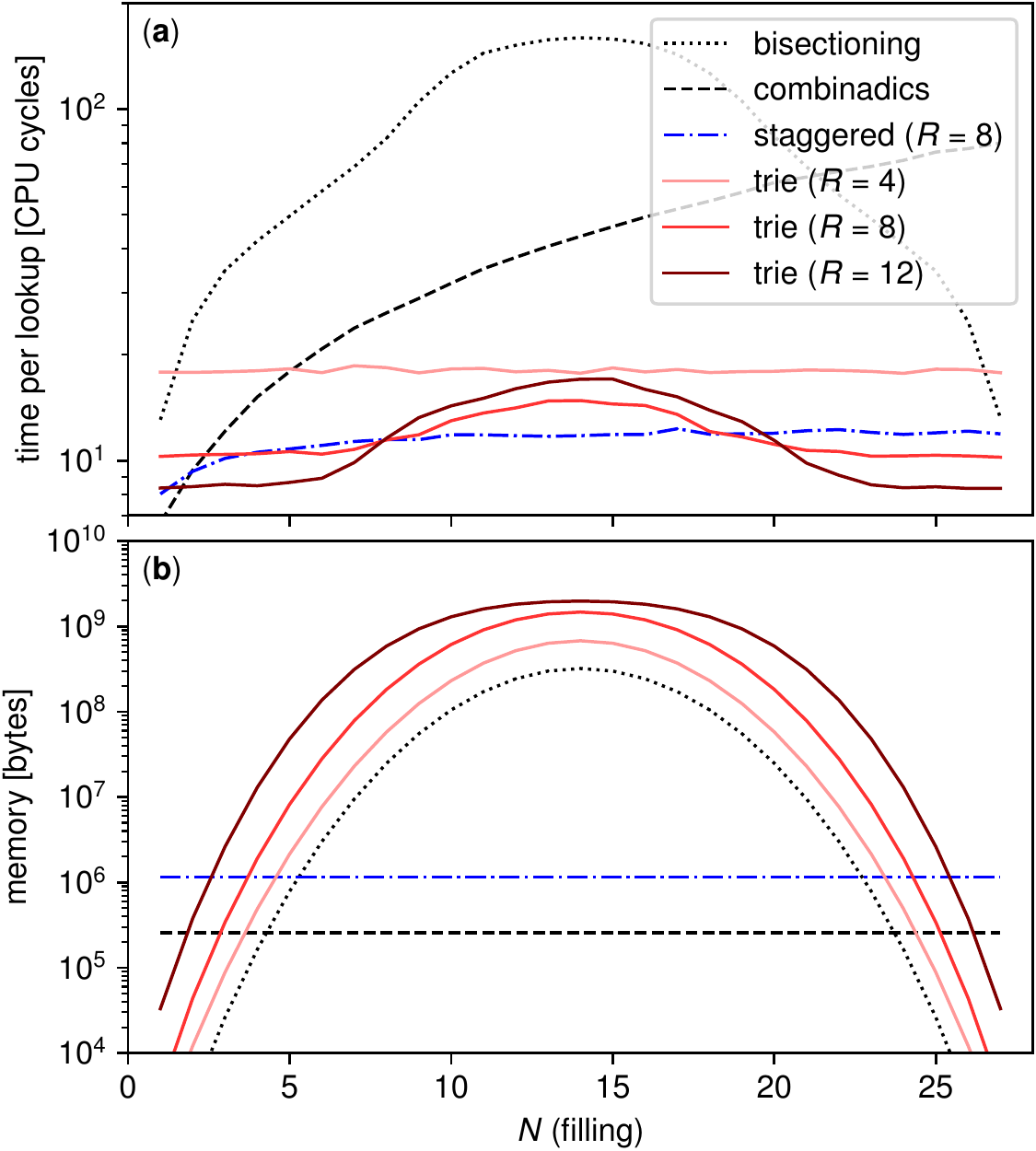}\caption{Lookup timings for combination with $M=28$ for sectors or varying
filling $N\in\{1,\ldots,27\}$.}

\label{fig:trie-timings}
\end{figure}

To this end, we implemented each ranking strategy---bisectioning,
combination ranking, staggered lookup and trie lookup---as Julia~\cite{Julia}
code and compiled with maximum optimization levels. We drew $10^{8}$
random states from each sector with replacement. We then ordered them
by numerical value to simulate the situation that while the states
generated in the application of Eq.~(\ref{eq:Talpha}) vary somewhat
unpredictably, they are usually still ordered to some degree (omitting
this ordering increases the cost of bisection search considerably
and unfairly.) We then timed 20 iterations on a single AMD Ryzen-3600 CPU core.

The results are presented in Fig.~\ref{fig:trie-timings}. Panel
(a) compares lookup times in terms of CPU cycles. We see that established
methods (bisectioning and combinadics ranking) are generally slowest,
except for the case of low filling: there, combinadics is most efficient
since its scaling is with $N$ rather than $M$. Otherwise, we observe
significant speedups for our improved algorithms, i.e., staggered and trie
lookups. This speedup is particularly pronounced for close to half-filled
sectors ($N\sim M/2$), since there the number of states is largest
and thus cache misses become more and more of an issue for bisectioning.
We further see that staggered lookup slightly outperforms tries for large
sectors for the same $R$, since it is more economical in terms of
cache demand. For a more sparsely populated sector (away from the
half-filled case), we see that tries are slightly more efficient.
In general, both trie and staggered lookups perform well, typically
requiring around 10--15 CPU cycles per lookup.

Fig.~\ref{fig:trie-timings}(b) presents the corresponding memory
demand for each index: here, the baseline is given by bisectioning,
which requires an ordered list of all states $N_s$ in the sector. The trie
indices add a radix-dependent overhead: for the half-filled case,
we find an overhead factor $f\approx2.11$ for $R=4$, $f\approx4.58$
for $R=8$, and $f\approx6.16$ for $R=12$. For the more sparsely
populated cases $N=5$, one finds that the overhead increases more
strongly: $f\approx2.73$, $10.45$, and $61.06$ for $R=4,8$, and
$12$, respectively. This is to be expected, as in the limit $R\to M$
we recover the inverse of the ``sparseness'' factor, $f\to2^{M}/N_{s}$.
Fortunately, the total memory demand of the index is still monotonously
falling as we increase sparseness.

\subsection{Hubbard chain}\label{sec:chain}

To examine the performance of ranking in a more realistic setting,
let us study a special case of Eq.~(\ref{eq:H}), namely a chain of
$M/2$ Hubbard atoms:
\begin{equation}
H=-t\sum_{i=2}^{M/2}\sum_{\sigma}[\cdag_{i\sigma}\cee_{(i-1)\sigma}+\mathrm{h.c.]}+U\sum_{i=1}^{M/2}\cdag_{i\uparrow}\cdag_{i\downarrow}\cee_{i\downarrow}\cee_{i\uparrow},\label{eq:Hring}
\end{equation}
where the sum over spins $\sigma$ runs over $\{\uparrow,\downarrow\}$.
We are interested in computing the lowest energy eigenvalue by
constructing the Krylov subspace
given sector $N,S_{z}$ (cf.~Eq.~\ref{eq:partrank})\footnote{Quantum numbers beyond $N,S_{z}$ have not been employed here since they cannot be implemented for the staggered combinadics, only for the tries.} using the Lanczos
algorithm, on-the-fly evaluation (\ref{eq:Talpha}). We implemented
these techniques in Julia and ran the calculations in parallel on
six AMD Ryzen-3600 CPU cores on the same node.

\begin{figure}
\includegraphics[width=1\columnwidth]{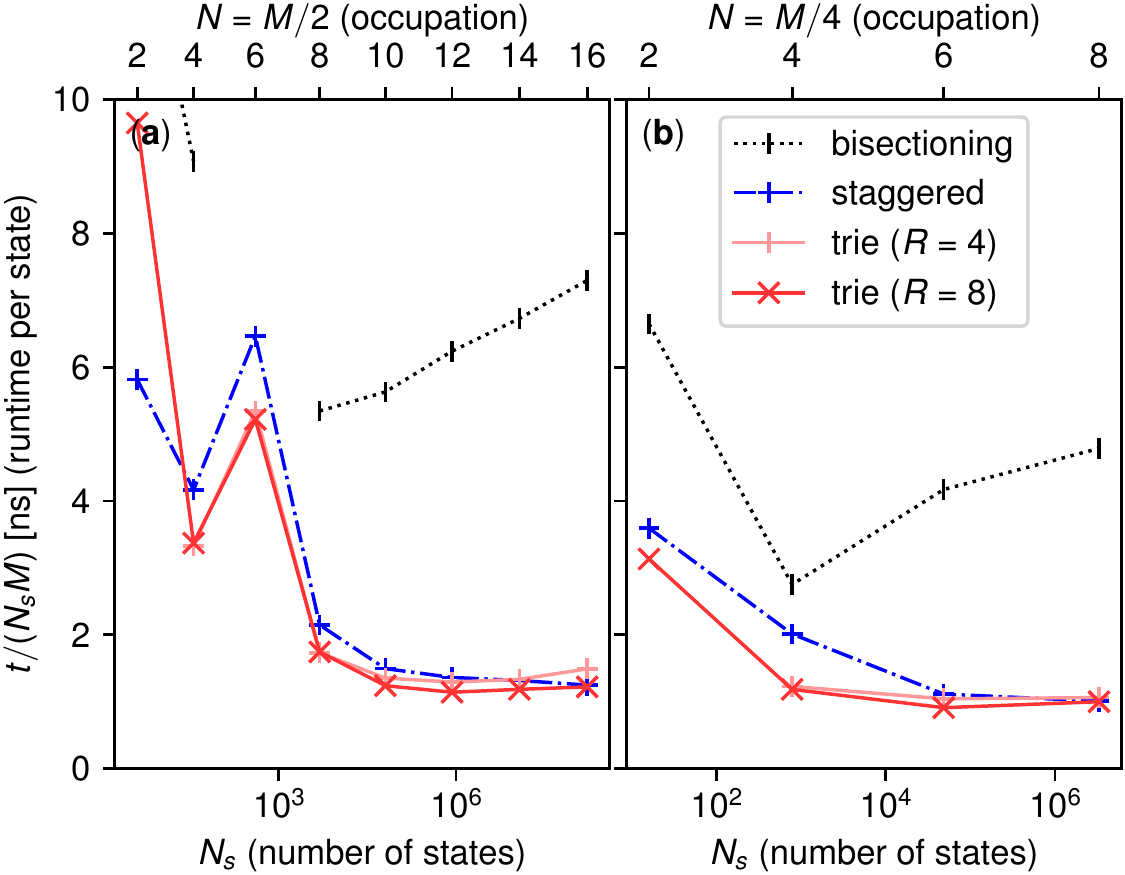}

\caption{Runtime of different ranking strategies when performing a vector--matrix
multiplication $H|\psi\rangle$ with (\ref{eq:Hring}) for two sectors: (a)
$N=M/2$ and $S_{z}=0$ for $M=4,8,\ldots,32$ and (b) $N=M/4$ and
$S_{z}=0$ for $M=8,16,24,32$. We plot the average runtime per state
over the number $N_{s}$ of states in the sector (bottom axis) and
the occupation $N$ (top axis), respectively.
}
\label{fig:eig-timings}
\end{figure}

Fig.~\ref{fig:eig-timings} presents the average runtime for a single
matrix--vector multiplication $H|\psi\rangle$ per state, i.e., the total runtime divided by the
number $N_{s}$ of states in the sector as well as divided by $M$, in units of nanoseconds.  We
see that both our ranking techniques, staggered and trie lookup, improve
substantially on the state-of-the-art bisectioning, achieving an around three-fold
speed-up.

We note that the timings in Fig.~\ref{fig:eig-timings} not only
include rankings. Indeed, Eq.~(\ref{eq:Talpha}) suggests the following,
rather crude, estimate for the total runtime:
\begin{equation}
t\approx N_{s}N_{t}(t_{\mathrm{rank}}+t_{\mathrm{apply}}+t_{\mathrm{FMA}}+t_{\mathrm{mem}}),\label{eq:t}
\end{equation}
where $N_{t}$ is the number of non-branching terms (in our case,
$N_{t}=3M/2$), $t_{\mathrm{rank}}$ is the time needed for ranking
states, $t_{\mathrm{apply}}$ is the time needed for computing the
bit operations when applying $T$ in Eq.~(\ref{eq:Talpha}), $t_{\mathrm{FMA}}$ is the
time needed for multiplying with the corresponding element of $\psi$
and accumulating the result, and $t_{\mathrm{mem}}$ is the time for
accessing the corresponding component of $\psi$. Disentangling the
constituent times is difficult, but profiling information suggests
that around 1/3 of runtime is spent on ranking in the case of staggered
and trie lookup. This is consistent with Fig.~\ref{fig:trie-timings}(a),
where we only observe a moderate increase of $t$ with $N$ even though
one expects linear scaling of $t_{\mathrm{rank}}$. In contrast, for
the bisectioning lookup the ranking times completely dominate the
computation.

Since trie and staggered lookup scale with $M$ rather than $N$,
the half-filled case $N=M/2$ in Fig.~\ref{fig:trie-timings}(a)
is most favorable for these algorithm. To study a more sparse setting,
 Fig.~\ref{fig:trie-timings}(b) presents timings for the quarter-filled case $N=M/4$. There,
bisectioning, which scales with $\min(N,M-N)$, cf.~Fig.~\ref{fig:trielookup},
improves compared to staggered and trie lookup. However, we still
observe an about four-fold speed-up by switching to the improved ranking
algorithms.

Let us finally note that the differences within the improved algorithms
are minor, consistent with the ranking microbenchmark in Sec.~\ref{sec:micro}.
Interestingly, even though trie lookup with $R=8$ consistently outperforms
$R=4$ in the microbenchmark, the two algorithms have similar performance
in the  benchmark Fig.~\ref{fig:trie-timings}. This may be due to the fact that
$R=8$ has a significantly larger memory footprints which becomes more relevant for the full algorithm, so the cache
misses may balance out the speed benefit.  (Cache pressure is expected to be higher in
this benchmark due to the manipulation of the state vectors.)

\subsection{Hubbard ring}

To showcase trie ranking (Sec.~\ref{sec:trie}), we choose a system and a set of quantum numbers where staggered lookup is not possible.  To do so, we consider again a one-dimensional system of
$M/2$ Hubbard atoms, but with periodic boundary conditions:
\begin{equation}
H = -2 t \sum_{k\sigma} \cos(k) \cdag_{k\sigma} \cee_{k\sigma}
+ U \sum_{kk'q} \cdag_{(k+q)\uparrow} \cdag_{(k'-q)\downarrow} \cee_{k\downarrow} \cee_{k'\uparrow},
\label{eq:Hk}
\end{equation}
where $\sigma\in\{\uparrow,\downarrow\}$, $k\in\{0, 4\tfrac\pi M,
8\tfrac\pi M, \ldots, (2M-2)\tfrac\pi M\}$ and we enforce periodic
boundary conditions by identifying $\cee_{k\sigma}\equiv \cee_{(k+2\pi)\sigma}$.

Let us note that $H$ in the momentum basis (\ref{eq:Hk}) is significantly less compact
than the real-space formulation, cf.~Eq.~(\ref{eq:Hring}), since the Hubbard interaction
generates $\bigO(M^3)$ terms.  For this reason, such symmetries are usually taken into
account implicitly in ED codes by generating a set of ``representative'' states in
the real-space basis and then symmetrizing the result to obtain the corresponding momentum
state~\cite{Lin90SublatticeCoding}.  However, there are other quantum numbers, such
as orbital parity in quantum impurity models (also known as the PS
quantum number~\cite{Parragh12PS}), which can be directly expressed in the real-space
occupation number basis.

These caveats nonwithstanding, the momentum basis admits a further quantum number, the
total crystal momentum:
\begin{equation}
    K_\mathrm{tot} = \sum_{k\sigma} k \cdag_{k\sigma} \cee_{k\sigma} \mod 2\pi,
    \label{Ktot}
\end{equation}
which commutes with both the Hamiltonian and the occupation number basis and can thus
be readily used to lower the block size.  Therefore, it provides a benchmark for trie
rankings vs. established bisectioning methods.

\begin{figure}
\includegraphics[width=1\columnwidth]{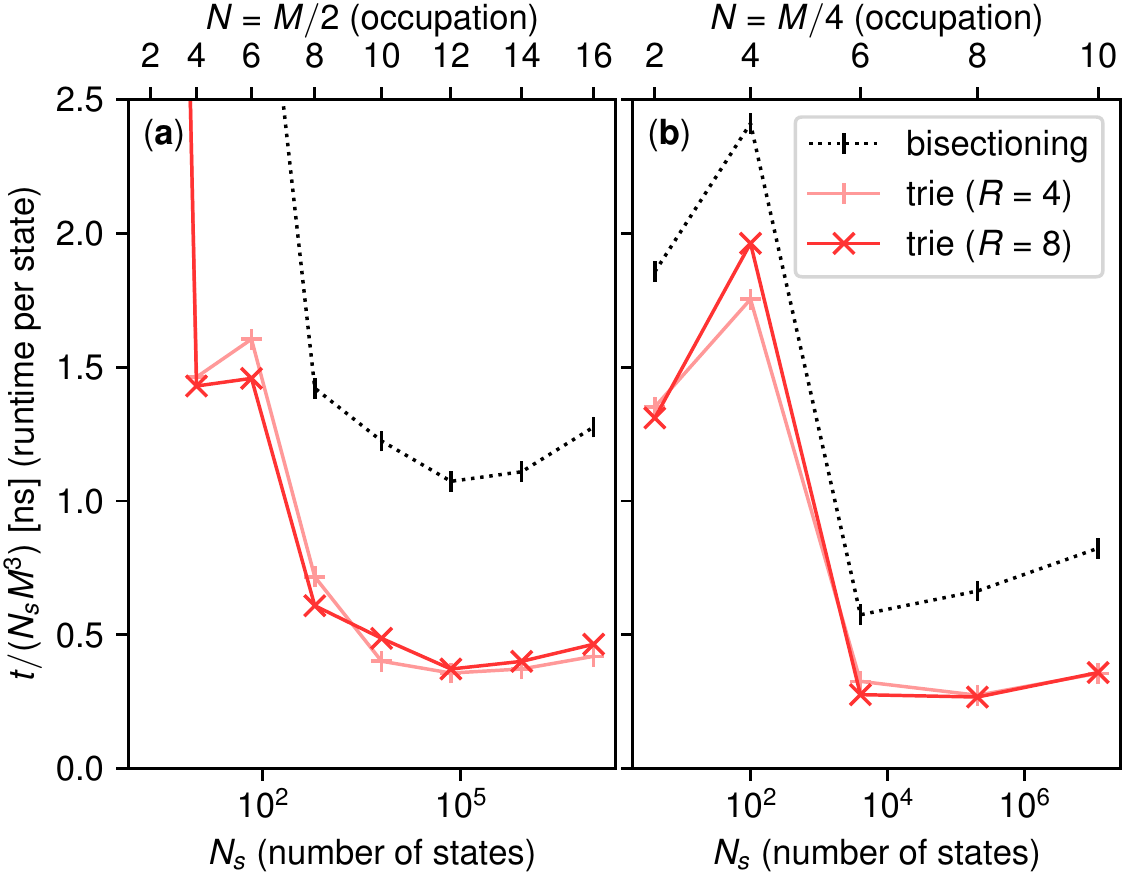}

\caption{Runtime of different ranking strategies when performing a vector--matrix
multiplication $H|\psi\rangle$ with (\ref{eq:Hk}) for two sectors: (a)
$N=M/2$, $S_{z,\mathrm{tot}}=0$, and $K_\mathrm{tot}=0$ for $M=4, 8, \ldots, 32$ and (b) $N=M/4$,
$S_{z}=0$ and $K_\mathrm{tot}=0$ for $M=8,16,\ldots,40$. We plot the average runtime per state
over the number $N_{s}$ of states in the sector (bottom axis) and
the occupation $N$ (top axis), respectively.
}
\label{fig:ring-timings}
\end{figure}

Fig.~\ref{fig:ring-timings} again presents the average runtime for a single
matrix--vector multiplication $H|\psi\rangle$ per state, i.e., the total runtime divided by the
number $N_{s}$ of states in units of nanoseconds.  Instead of by $M^{-1}$ as in Fig.~\ref{fig:eig-timings}, the runtime is scaled
with $M^{-3}$, since that is the scaling of the number of terms in Eq.~(\ref{eq:Hk}).
We see that the speed-up of trie lookup vs. bisection is milder here than in
Sec.~\ref{sec:chain}. We can attribute this to the greater ``sparseness'' of states
in the space of $M$-bits, which means that tries have to process more bits per
state.  However, the benefit is still substantial, with a two- to three-fold
improvement in runtime observed.

Notably, instead of plateau for trie-based ranking in Fig.~\ref{fig:eig-timings},
we see an uptick in runtime for large $M$ in Fig.~\ref{fig:ring-timings}.
This may be indicative of growing cache pressure in this cases due to the memory
footprint of the trie.

\section{Conclusions and outlook}

We have removed a computational bottleneck of exact many-fermion calculations
by speeding up the ranking of states, thereby providing an efficient
mapping between indices of state in the full Fock space and in the
block corresponding to a set of quantum numbers. For the common case
of conserved particle number, the state-of-the-art combinadics algorithm
present in many ED algorithms can be modified to yield considerable ranking
speed-ups at negligible overhead with a {\em staggered lookup}. For more complictated quantum number
combinations, the state-of-the-art bisectioning algorithm can be replaced by a {\em trie lookup},
which has a similar scaling with memory but a much-improved performance.
Thanks to these improvements, ranking is no longer the bottleneck of the ED code.

With these optimizations, we expect exact diagonalization to stay
competitive as, e.g., a solver for quantum impurity models~\cite{DMFT}
in cases where the quantum Monte Carlo techniques encounter a significant
sign problem or where high numerical accuracy is required for, e.g.,
analytic continuation~\cite{Otsuki17,Nevanlinna}.

Trie ranking is also applicable to non-Abelian symmetries~\cite{Lin90SublatticeCoding}
leveraged in spin systems, where instead of ranking states in a quantum
number sector we are ranking representatives of a symmetry orbit within
a symmetry sector. Combining trie lookup and sublattice coding techniques~\cite{Lin90SublatticeCoding,Lauchli18}
seems particularly promising, as sublattice coding reduces the size
and scaling of the states which have to be ranked, which should make
the memory overhead associated with tries less problematic also for
large systems.

An intriguing challenge for trie rankings are selected CI techniques~\cite{Huron73} as well as related Monte Carlo approaches~\cite{FCIQMC,Jaklic94},
which restrict FullCI to a subset of states in the Fock space. These
techniques optimize the set, either stochastically or deterministically,
by minimizing the ground state energy, which means the set to be ranked
cannot be static. This ``dynamic ranking'' makes the use of linearized
and packed tries impractical, so ranking based on hash tables could
be competitive.

A unit-tested Julia implementation of the techniques outlined here
is available from the authors upon reasonable request and forthcoming
as an open-source package.

\section*{Acknowledgements}

We thank A. Kauch, F. Krien, H. Hofst\"atter, and C. Watzenb\"ock for
fruitful discussions and careful review of the manuscript, and acknowledge
supported by the FWF (Austrian Science Funds) through projects P30819
and P32044.

\appendix

\section{Non-branching term rules}
\label{app:mul}

For completeness, we show here how to construct general rules for non-branching
terms and arbitrary products thereof.  Most of this material is well-known, but
the product rule has to the best of our knowlege not been stated in this general form.

We begin by restating Eq.~(\ref{eq:Talpha}) in a slightly altered form: let $T$ be an
arbitrary product of creation and annihilation operators (not necessarily
normal-ordered) with a scalar prefactor.  Then there exists a tuple ($v,m,l,r,x,s$) of numbers such that the application
of $T$ on any basis state $|\alpha\rangle$ in the occupation number basis is:
\begin{align}
T|\alpha\rangle&=v\delta_{\alpha\owedge m,r}(-1)^{\popcount(\alpha\owedge s)}|\alpha\oplus x\rangle,\label{eq:Talpha1}\\
\langle\alpha|T&=v\delta_{\alpha\owedge m,l}(-1)^{\popcount(\alpha\owedge s)}
\langle\alpha\oplus x|,\label{eq:Talpha2}
\end{align}
where $\owedge$ denotes bitwise \bitand, $\oplus$ denotes bitwise
\bitxor, and $h$ is the Hamming weight, i.e., the number of set bits.
(We have added an element $l$ for later convenience.)

By way of a proof, we will offer an algorithm to construct $(v,m,l,r,x,s)$.  Let
us start with the ``building blocks'', the annihilation operators $\cee_i$.  By
comparing Eq.~(\ref{eq:Talpha1}) with Eq.~(\ref{eq:cee}), one finds that:
\begin{subequations}%
\begin{align}
   v &= 1, & m &= 2^i, \\
   l &= 0, & r &= 2^i, \\
   x &= 2^i, & s &= 2^i - 1.
\end{align}%
\label{eqs:ci}%
\end{subequations}
The values for corresponding creation operator $\cdag_i$ are obtained by simply exchanging $l$ and $r$.
(The same is true for the transpose of any other term.)

\begin{figure}
\begin{algorithmic}[1]
\Function{{\rm $t = $} mul}{$t_a, t_b$}
    \State $(v_a, m_a, l_a, r_a, x_a, s_a) \gets t_a$
    \State $(v_b, m_b, l_b, r_b, x_b, s_b) \gets t_b$

    \If{$(r_a \oplus l_b) \owedge m_a \owedge m_b \ne 0$}
         \State $v\gets 0$
    \Else
         \State $v\gets v_a v_b$
    \EndIf
    \State $m \gets m_a \ovee m_b$
    \State $r \gets r_b \ovee (r_a \owedge \neg m_b)$
    \State $l \gets l_a \ovee (l_b \owedge \neg m_a)$
    \State $x \gets l \oplus r$
    \State $s \gets s_a \oplus s_b$
    \State $s \gets s \owedge \neg m$
    \State $z \gets r$
    \State $p \gets z \owedge s_b$
    \State $z \gets z \oplus x_b$
    \State $p \gets p \oplus (z \owedge s_a)$
    \State $v \gets v (-1)^{h(p)}$
    \State $t \gets (v, m, l, r, x, s)$
\EndFunction
\end{algorithmic}
\caption{Fast algorithm for constructing the tuple $t$ for on-the-fly multiplication
for a product $T=T_aT_b$, where the factors $T_a$ and $T_b$ are encoded by tuples $t_a$ and $t_b$, respectively.
Here, $\owedge$ denotes bitwise \bitand, $\ovee$ denotes bitwise \bitor, $\oplus$ denotes
bitwise \bitxor, $\neg$ denotes bitwise \bitnot, and $h(x)$ counts the number of set bits
in $x$.
}
\label{fig:mul}
\end{figure}

We can now understand the role of each of the elements in the tuple: $m$ is the bitmask, with bit $i$ set whenever there is any operator in $T$ with flavor $i$.
$r$ is the ``demands'' of $T$ on its right side, i.e., for any flavor in $m$, the state $\alpha$ must have the same occupation as $r$ does whenever $T$ is applied from the left or the result will be zero.  The same holds for $l$ when $T$ is applied from
the right.  The tuple $(v,m,l,r)$ is enough to encode the full term ($m$ can be omitted if normal-ordering is imposed), but for performance reasons it is advantageous to store two more fields:  set bits in $x$ correspond to those flavors which will be changed by $T$.  Finally, $s$ is the sign mask, which encodes the anticommutativity rules: whenever $T$ is applied to $\alpha$ any flavor in $\alpha$ that is also in $s$ will cause the result to flip sign.

To complete the proof, we need to construct the product $T=T_aT_b$ of two
terms.  We present this algorithm in Fig.~\ref{fig:mul}.  There, $t = (v,m,l,r,x,s)$ is
the desired result, tuple for $T$.  The inputs are the tuple $t_a = (v_a, m_a, l_a, r_a, x_a, s_a)$
for $T_a$ and $t_b = (v_b, m_b, l_b, r_b, x_b, s_b)$ for $T_b$.

Lines 4--8 check the Pauli principle.  This is relatively easy: let us denote by $|\psi\rangle$ the
``intermediate'' state in the application of $T$, i.e., $T|\alpha\rangle=T_a|\psi\rangle$.
$r_a$ and $l_b$ are then ``demands'' of $T_a$ and $T_b$, respectively, on $\psi$, each
each confined to their mask.  This means we simply have to fulfill both demands whenever the masks
intersect $m_a \owedge m_b$.  When the demands are not fulfilled, the Pauli principle is violated
and the value $v$ of the term is set to zero (line 7), otherwise it is---for now---simply the product
of both scalars (line 5).

Line 9--14 deal with the actual composition of the two terms. Firstly, in line 9, we construct the
mask $m$ as union of the individual masks, since the operator flavors in the product are just the
union of the operators in each factor.  For the right outer state $r$, $r_b$ takes precedence whenever
$m_b$ is set, otherwise, the other term can make demands $r_a$ (line 10).  For the left outer state $l$,
we make a similar argument (line 11).  The change mask $x$ is given simply as the symmetric bit
difference between left and right (line 12).

The sign mask $s$ is at first set as the simple exclusive or of the individual masks (line 13), however, there is a complication: the creation and annihilation operators alter the state $\alpha$, so each flavor in $m$
is ``locked'' in place by $T$.  This means we need to exclude any flavor from the sign mask $s$ that is
also present in $m$ (line 14).

The term $T'$ we have constructed so far is almost equal to $T=T_aT_b$, however, it may still deviate from
$T$ by a sign due to the reordering of creators and annihilators encoded in the tuple $t$. Instead of
keeping track of these permutations explicitly, we use a computationally appealing shortcut: we simply
compare the effects of $T_aT_b|\alpha\rangle$ and $T'|\alpha\rangle$ for a state where $T|\alpha\rangle\ne 0$.
One such state is, by construction, $|r\rangle$.  Lines 15--18 compute Eq.~(\ref{eq:Talpha1}) for $T_aT_b|r\rangle$ and keep track of the relevant sign masks overlaps in $p$.  Any deviation between $T$ and $T'$
is then absorbed into $v$ (line 19) and the algorithm is complete.

We note that the algorithm in Fig.~\ref{fig:mul} does not only prove that a tuple for on-the-fly application
exists, it also offers an extremely fast way to compute products of terms on modern machines, as it
relies exclusively on bit operations.

\end{document}